# Spring-block friction model for landslides: Application to Vaiont and Maoxian landslides

Rong Qiang Wei[ab], Qing Li Zeng[ab]

a University of Chinese Academy of Sciences, Beijing 100049, China

b Key Laboratory of Computational Geodynamics, Chinese Academy of Sciences, Beijing 100049, China

**Abstract**

It is necessary to study the kinematics of the landslide prior to its failure for estimating the time of landslide instability. Based on a spring-block friction model, considering the Dieterich-Ruina's friction law, the kinematic displacement and velocity of landslide along the slip surface are analyzed under quasi-static approximation. A three-parameter algebraic relationship between the displacement (or velocity) and time is obtained, and then applied to two typical landslides: Vaiont in Italy, and Maoxian in China. The results show that this spring-block friction model can well describe the kinematic data of landslides before their failure. If the effective data of displacement can be obtained to determine the three parameters above, this simple physical model could be used to estimate the time of landslide instability. This spring-block friction model also provides clear physical basis for the usual inverse-velocity method of the landslide warning, the stick-slip of some landslides, and the scaling relationship between the numbers of the landslides and their volume.

## 1. Introduction

Catastrophic landslides often cause serious personnel and property damage without the significant precursors. Detailed and continuous field observations show, however, that landslide failure is a progressive process that can last from days to decades (eg., Saito and Uezawa, 1961; Voight, 1988). Slope deformation is the important stage before the landslide failure, and the corresponding displacement and velocity are the typical and abundant monitoring physical quantities. The typical (cumulative) displacement-time or velocity-time curve (Fig. 1) of the slope deformation shows that the displacement or velocity of the slope along the slip surface increases slowly at the beginning, but enlarges sharply when the slope is approaching failure. Many authors have studied these displacement-time or velocity-time curves of landslide and put forward a lot of methods so as to predict the occurrence of landslide (eg., Saito, 1969; Kilburn and Petley, 2003; Carlà et al., 2017).

The variation of velocity with time prior to the landslide failure can be considered as the accelerated creep stage in the tertiary creep. Based on this assumption and the experiments, some authors have given the empirical relationship between the surface acceleration and the surface velocity of the landslide (eg., Saito and Uezawa, 1961; Saito, 1969; Kennedy and Niermeyer, 1971; Kilburn and Petley, 2003), ie.,

$$\frac{dv}{dt} = Av^\alpha \qquad (1)$$

where $\alpha$ and $A$ are constants, $v$ the velocity, $t$ the time.

From Eq. (1), one can obtain (2) (eg., Bhandari, 1988),

$$t_c - t \sim (\frac{1}{v})^{\frac{1}{\alpha-1}} \qquad (2)$$

where $t_c$ is the landslide failure time.

Expression (2) is consistent with the observations by Voight (1988) if $\alpha = 2$, that is, the slip velocity $v \sim 1/(t_c - t)$. This relationship is the basis for the inverse-velocity method of landslide warning (eg., Rose and Hungr, 2007; Wang et al., 2015; Carlà et al., 2017).

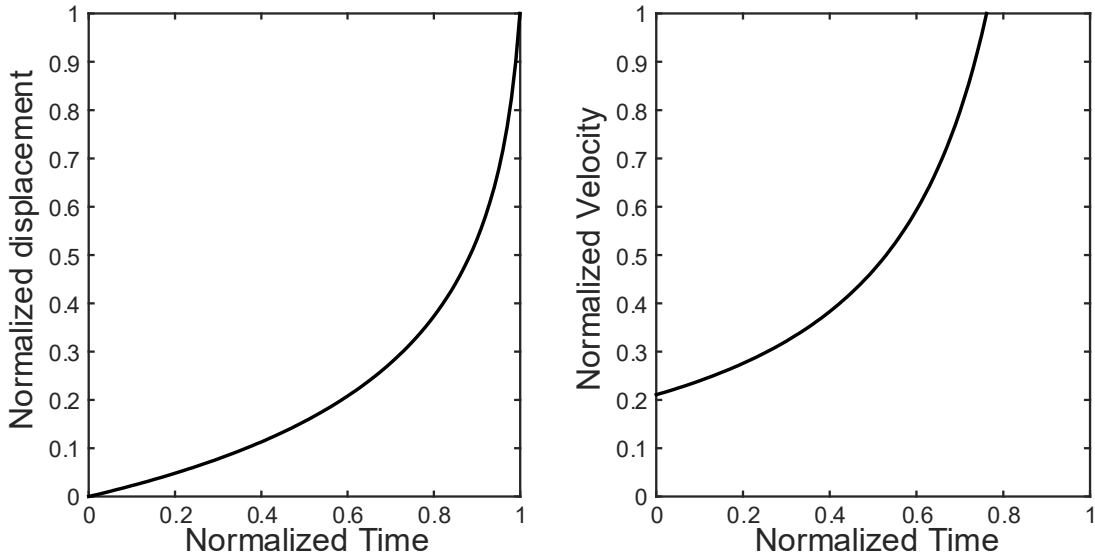

Fig. 1 Typical (cumulative) curve for the normalized displacement-time (left) or velocity-time (right) of the slope deformation before landslide failure. Normalized displacement and velocity are calculated with Eq. (8) and (10), respectively.

In order to deal with the above phenomena, Helmstetter et al. (2004) developed a slider block model, in which the Dieterich-Ruina friction law is taken into account. Their work provided a clear physical rather than empirical basis for analysing of the velocity-time relationship before the landslide failure.

Here we develop a spring-block friction model to study the kinematic displacement-time and velocity-time before landslide failure using the Dieterich-Ruina friction law, mainly base on the work of Popov et al. (2012). In this model the traction is definitely taken into account. A simple three-parameter algebraic formula is presented to analyze the displacement-time or velocity-time observations, and applied to the famous Vaiont landslide in Italy, and Maoxian landslide in China. Finally, the possibility of landslide warning using observed displacement data is discussed.

## 2. Spring-block model with Dieterich-Ruina friction law

As stated above, there is a slow developing processes before the landslide failure. This is similar to the earthquake nucleation. Referring to Popov et al. (2012) for the study on earthquake nucleation processes, we establish a simple physical model of a frictional pair as shown in Fig. 2a to study the "nucleation" of a landslide. In this frictional pair, one of the partners is modeled as a rigid slope and the other as a rigid block of mass $m$. The total elasticity of the system is represented as a spring with the stiffness $c$. The block is loaded over a rigid slope through the gravitational force and the spring. The spring is extended at a load point velocity $v_0$. The friction coefficient between the slope and the block is defined by the Dieterich-Ruina friction law (Eq. (3)).

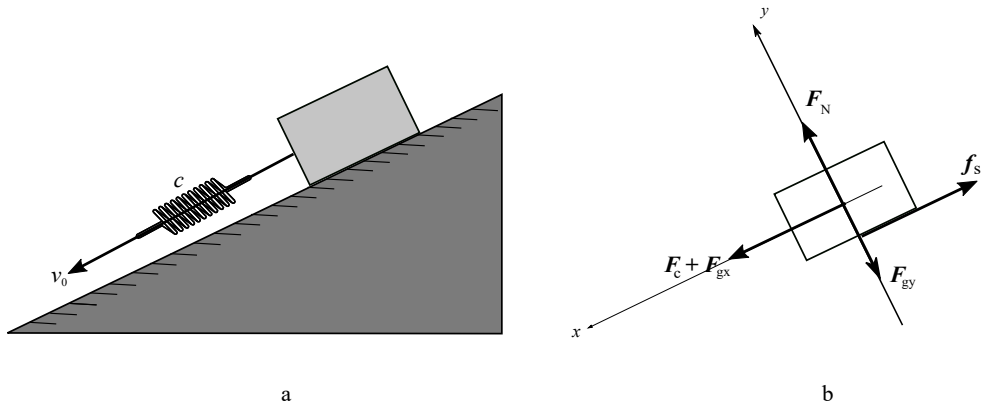

Fig. 2 (a) The sketch for the spring-block friction model; (b) Force analysis for the block. In the x-direction, the block is loaded by the coaction of the spring force $F_c$, and the component of the gravitational force $F_{gx}$, and the friction $f_s$. In the y-direction, the block is in equilibrium by the coaction of the support force $F_N$ exerted by the slope and the component of the gravitational force $F_{gy}$.

Dieterich-Ruina friction law states friction depends both on the instantaneous sliding velocity $v$ and a time-dependent state variable $\theta$ (Ruina, 1983),

$$\begin{cases} \mu = \mu(v,\theta) = \mu_0 + a\ln(\frac{v}{v^*}) + b\ln(\frac{v^*\theta}{D_c}) \\ \frac{d\theta}{dt} = 1 - \frac{v\theta}{D_c} \end{cases} \quad (3)$$

where $\mu$, $\mu_0$ are friction coefficient and that at some steady state reference velocity $v^*$, respectively. $a$, $b$ are both positive constants, and $D_c$ the critical slip distance. This friction law can apply not only to rock, but also to a wide variety of materials such as soil, plastic and glass.

The slowly developing stage of the landslide failure can be treated as quasi-static (zero acceleration), ie., equilibrium conditions is fulfilled for every point in time. According to Fig. 2b, in the direction of the movement of the block, the equilibrium condition is $F_{gx} + F_c = f_s$, namely,

$$F_{gx} + c(x_0 + v_0 t - x) = \mu(v, \theta) F_N \quad (4)$$

where $x_0$ is the extension of the spring when the slip of the block occurs. It should be pointed out $F_{gx}$ and $F_N$ here are not expressed explicitly, which are related to the mass of the block and the inclination of the slope, and they will be shown in the subsection 5.3.

If the parameters in Eq. (3), (4) and the related initial conditions are known, the slip displacement and the velocity of the block can be calculated. However, these equations are nonlinear and generally can only be numerically solved. Popov et al. (2012) showed some numerical examples in which $F_{gx}$ was not taken into account.

## 3. An analytic solution of spring-block friction model

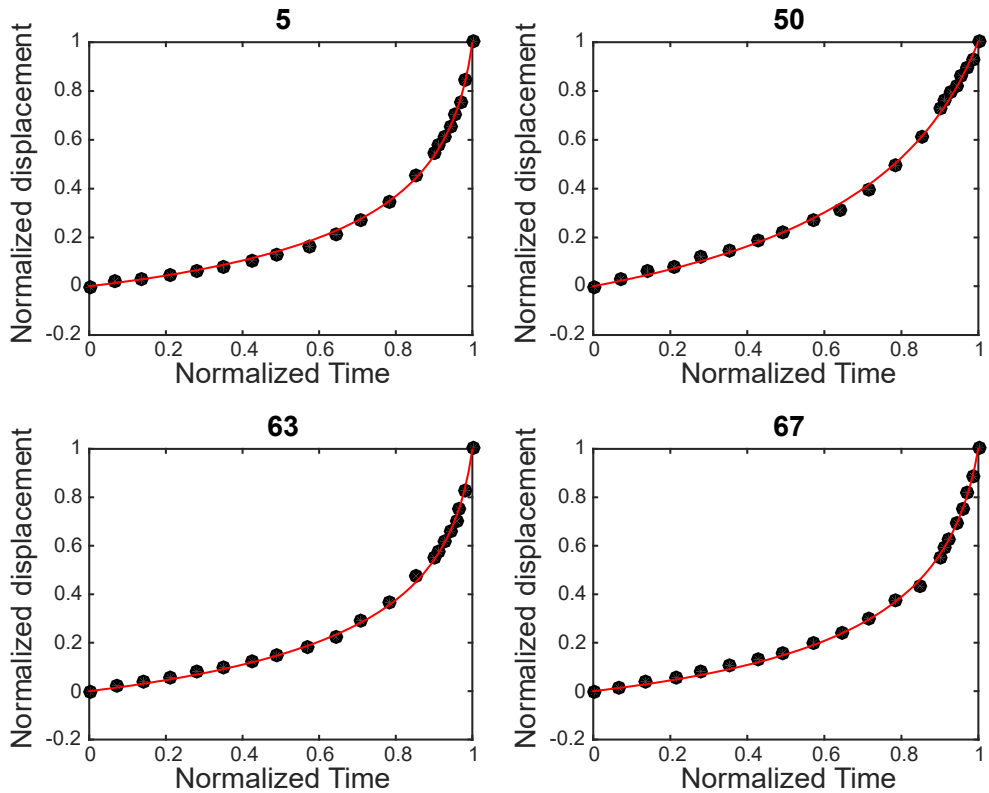

Fig. 3 Normalized displacement measurements vary with time for four benchmarks of the Vaiont landslide (black dots, digitized from Helmstetter et al. (2004)). The red lines are fitted with the Eq. (8).

For the kinematics of spring-block model of the landslide, the condition that $v\theta/D_c \gg 1$ holds (eg., Helmstetter et al., 2004; Popov et al., 2012).

Thereby, the second equation of (3) can be reduced to,

$$\frac{d\theta}{dx} = -\frac{\theta}{D_c}$$

and one can obtain $\theta$ immediately,

$$\theta = \theta_0 \exp(-\frac{x}{D_c}) \tag{5}$$

Combining the first equation in (3), Eq. (4) and (5), one can obtain,

$$F_{gx} + c(x_0 + v_0 t - x) = F_N[\mu_0 + a \ln(\frac{dx/dt}{v^*}) + b \ln(\frac{\theta_0 v^*}{D_c}) - \frac{bx}{D_c}] \tag{6}$$

And one can get further,

$$\frac{dx}{dt} = A \exp(\frac{B}{a}x + \gamma t) \tag{7}$$

where,

$$A = v^* \exp\left(\frac{-\mu_0 - b \ln \frac{\theta_0 v^*}{D_c} + \frac{F_{gx} + cx_0}{F_N}}{a}\right)$$

$$B = \frac{b}{D_c} - \frac{c}{F_N}, \gamma = \frac{cv_0}{aF_N}$$

Finally, the displacement $x$ can be obtained from equation (7) with initial conditions: $x = 0$ when $t = 0$. That is,

$$x = -\alpha \ln[1 - \beta(\exp(\gamma t) - 1)] \tag{8}$$

where $\alpha = \frac{a}{B}, \beta = \frac{AB}{a\gamma}$

The time for landslide failure is calculated using the condition that the argument of the logarithm in Eq. (8) is zero:

$$t_c = \frac{1}{\gamma} \ln(1 + \frac{1}{\beta}) \tag{9}$$

Accordingly, the slip velocity of the block is,

$$v = \frac{\alpha\beta\gamma \exp(\gamma t)}{1 - \beta[\exp(\gamma t) - 1]} \tag{10}$$

It should be noted that generally the block and the slope are not rigid, the Eq. (6) should be expressed by the means of "stress". However, the form of the Eq. (6) is almost unchanged.

Our model has clear improvements over the work by Helmstetter et al. (2004) and Popov et al. (2012): Firstly, we explicitly consider the effect of traction of the spring and the gravitational force, which is not taken into account by Helmstetter et al. (2004). Secondly, the displacement-time or velocity-time relationship we obtained is a simple algebraic equation rather than a system of differential equations by Helmstetter et al. (2004). Thirdly, we consider the block model on the slope and the traction of gravitational force on the block, while Popov et al. (2012) consider the spring-block model on the horizontal plane. Finally, we combine the parameters in Eq. (8) and (10), which are not easy to be obtained or measured, into three parameters. It will facilitate to analyze the monitoring data of displacement or velocity.

## 4. Case applications

The above expressions for displacement and velocity (Eq. (8) and (10)) include many parameters. In general, these parameters are difficult to be obtained or measured, and they may not be the same at different representative points for the same landslide. Therefore, in this section, we mainly use the functional forms of Eq. (8) and (10), and treat $\alpha$, $\beta$, $\gamma$ as characteristic parameters. These three parameters will be determined empirically with the measured data of slip cumulative displacement or velocity. Based on this, we discuss the possibility of estimating the landslide failure time, the unknown slip displacement or velocity. Here two typical landslides, ie., the Vaiont landslide of in Italy, and Maoxian landslide in China are taken as two examples.

### 4.1 Data collection and pretreatments for cumulative displacement

The world-famous Vaiont landslide occurred in Italy in 1963. There are a lot of studies on its developing processes, properties and mechanism. Multi-physical data, including slip displacement, have been monitored for many years before the landslide failure (eg., Genevois & Ghirotti, 2005). Displacement and velocity data used for Vaiont landslide in this paper are digitized from Helmstetter et al. (2004), in which the data for the benchmark points of No. 5, 50, 63 and 67 are digitized. For the convenience of studying and partially reduction of the error caused by digitization and other reasons, both the time and displacement are normalized to their maximum so that their values are in the range of [0, 1]. These pretreatments on the monitoring data are also used for the landslide of Maoxian.

The Maoxian landslide occurred on June 24, 2017, and is located in Maoxian County, Sichuan Province, China. Details on this landslide can be referred to (Zeng et al., 2018; Intrieri et al., 2017). Displacement data are digitized from Intrieri et al.

(2017), in which displacement data from Interferometric Synthetic Aperture Radar (InSAR) analysis for four points of Point 1, Point 2, Point 3, and TS are used.

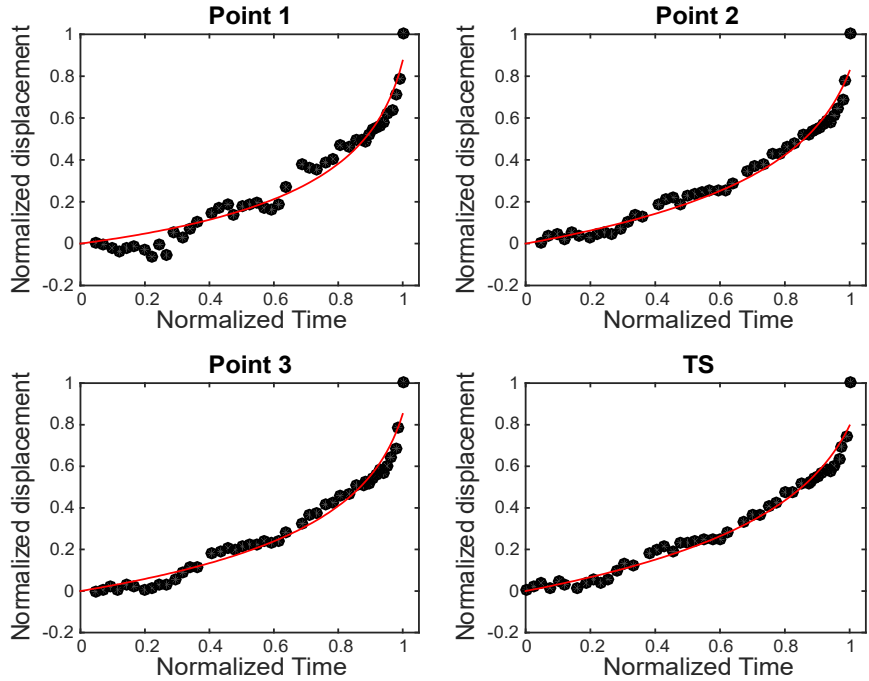

Fig. 4. Normalized displacement measurements varying with time for four points of the Maoxian landslide (black dots, digitized from Intrieri et al., (2017). The red lines are fitted with the Eq. (8).

### 4.2 Analysis of the cumulative displacement with spring-block friction model

We use the function form of Eq. (8) to fit the above preprocessed displacement data to obtain the characteristic parameters $\alpha$, $\beta$, $\gamma$. Here $\beta > 0$ and $\gamma > 0$, which can be inferred from the actual parameters they contain. Therefore, the fitting here is in a sense of constrained least-squares. And because Eq. (8) is non-linear, we use a genetic algorithm to produce relatively good initial values.

Fig. 3-4 show the fitting results for the landslides of Vaiont and Maoxian, respectively. Table. 1 lists the fitting values of the $\alpha$, $\beta$, $\gamma$. It can be seen that the curves fit well the observed normalized displacement data, indicating that the displacement of the landslide before failure can be well represented by the spring-block friction model. Therefore, this model is observationally valid and the quasi-static assumption holds. If the characteristic parameters $\alpha$, $\beta$, $\gamma$ can be accurately obtained, the displacement before landslide failure can be completely estimated by the spring-block friction model here. In this way, the physical process of landslide before failure can be accurately estimated.

### 4.3 Calculation of the slip velocity for the Vaiont landslide

Since the characteristic parameters $\alpha$, $\beta$, $\gamma$ can be obtained by fitting the cumulative displacement, we can further calculate the slip velocity of the landslide varying with time before failure with Eq. (10). Because only the velocity data of the Vaiont landslide is collected, we report the results of this landslide as an example. Fig. 5 shows the comparison of the normalized velocities observed (black dots) with those calculated with Eq. (10) (red line), in which $\alpha$, $\beta$, $\gamma$ are taken from Table.1. It can be seen that the velocities calculated are consistent with the observations. The good consistence indicates that the deformation velocity of the landslide before failure can be well represented by the spring-block friction model if the characteristic parameters $\alpha$, $\beta$, $\gamma$ can be accurately obtained. These show again that the physical process of landslide before failure can be accurately estimated by this physical model.

**Table.1 Characteristic parameters and the time for failure of Vainot and Maoxian landslides**

| Monitoring points | $\alpha$ | $\beta$ | $\gamma$ | $t_c$ | $\alpha^*$ | $\beta^*$ | $\gamma^*$ | $t_c^*$ |
|---|---|---|---|---|---|---|---|---|
| Vainot 5 | 0.31 | 0.71 | 0.86 | 72(70) | 0.64 | 0.29 | 1.33 | 72 |
| Vainot 50 | 0.59 | 0.56 | 0.90 | 80 | 0.50 | 2.76 | 0.27 | 74 |
| Vainot 63 | 0.31 | 0.83 | 0.77 | 73 | 0.53 | 0.95 | 0.64 | 72 |
| Vainot 67 | 0.36 | 0.49 | 1.07 | 73 | 0.49 | 1.56 | 0.44 | 73 |
| Maoxian Point1 | 0.33 | 0.83 | 0.75 | 1092(1040) | 18.42 | 0.01 | 2.61 | 1436 |
| Maoxian Point2 | 0.32 | 8.14 | 0.11 | 1091 | 1.34 | 0.70 | 0.56 | 1288 |
| Maoxian Point3 | 0.31 | 3.84 | 0.22 | 1090 | 1.50 | 0.28 | 1.02 | 1210 |
| Maoxian TS | 0.35 | 12.54 | 0.07 | 1078 | 3.24 | 0.22 | 0.79 | 1654 |

Notes: 1. Characteristic parameters with * in the 6-9$^{th}$ columns are the results from part monitoring data, in which the first 15 data for Vainot landslide, and 30 for Maoxian, are used. 2. $\alpha$ should be in m, and $\gamma$ should be in s$^{-1}$. For the normalized displacement and normalized time here, they are the dimensionless. $\beta$ is dimensionless. $t_c$ is in d. 3. The number in the square brackets is the actual days for the landslide failure.

It should be noted that the slip velocities here are normalized by $x_{max}/t_{max}$ and the time is normalized by $t_{max}$ ($x_{max}$ is the maximum displacement, $t_{max}$ the corresponding time), because $\alpha$, $\beta$, $\gamma$ are obtained by fitting the normalized displacement ($x/x_{max}$) and the normalized time ($t/t_{max}$).

### 4.4 Possibility for estimating the landslide failure time

It is very interesting whether it is possible to estimate the time of landslide failure using monitoring displacement data. According to the previous results, we can try this using Eq. (9) if $\beta$ and $\gamma$ are obtained by fitting the monitoring displacement data. Table. 1 shows some examples. It can be seen that the estimated failure time is generally delayed. The Vainot is delayed by 2-10 days, and Maoxian 38-52 days. The reason is obvious: The time in Eq. (9) is that when the displacement $x = +\infty$, but the actual displacement can not reach infinity. Therefore the time predicted by Eq. (9) should always be delayed.

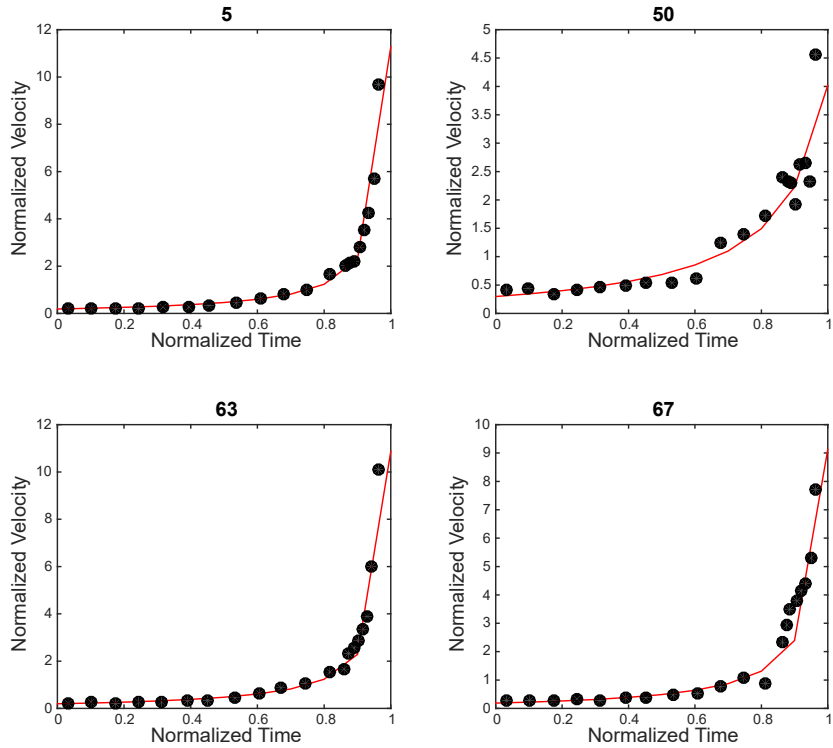

Fig. 5 Observations of normalized slip velocity-normalized time (black dots, the original data are digitized from Helmstetter et al. (2004)) for four benchmarks of the Vaiont landslide. The red lines are calculated with the Eq. (10). Details of normalization can be seen in the text.

Table. 1 also lists the estimated time for landslide failure only by using partial displacement data. It can be seen that the failure time is also delayed. This is understandable because the resulting displacement-time curve is relatively flat, corresponding to a long time of landslide failure. For the Vainot landslide, the failure time estimated by the partial data is only delayed 2-3 days. However, this may be an illusion, probably because the amount of the monitoring data is small, and the monitoring displacement is too good (smooth).

Therefore, if we want to use the monitoring displacement curve to estimate the time of landslide failure, we should pay attention to three aspects at least: (1) How to choose a reasonable and representative monitoring point? Generally we may use the data at sevral monitoring points to get a mean failure time; (2) The amount of the high-precision monitoring data should be as large as possible, especially those data close the landslide failure; (3) How to obtain reasonable characteristic parameters $\beta$, $\gamma$? This demands excellent nonlinear algorithms. In recent years, a large amount of satellite deformation data (such as InSAR, ground-based radar, etc.) have been accumulated. On the other hand, there are many good data processing algorithms. All of these make it possible to timely warn the landslides failure.

# 5. Discussions

## 5.1 Slip velocity near the landslide instability

Near $t_c$, since $\exp(\gamma t) = \exp(\gamma t_c) + \gamma \exp(\gamma t_c)(t - t_c) + ...$, Eq. (8) can be approximated to,

$$x \approx -\alpha \ln[\beta\gamma(1 + \frac{1}{\beta})(t_c - t)] \tag{11}$$

Further, one can obtain the velocity as,

$$v \approx \frac{\alpha}{t_c - t} \tag{12}$$

Eq. (12) is just the observations of Voight (1988), and the basis for the inverse-velocity method and the like. Since Eq. (12) is true only near $t_c$, the inverse-velocity and similar method can only be applied to velocity data near landslide failure. For the whole set of monitoring velocity data including those far from the landslide failure, $v \sim t$ is not a simple reciprocal relationship. This requires paying attention when inverse- velocity method and the like are applied.

## 5.2 About the effects of fluids

The effects of the fluids includes the physical effect and the chemical effect. Neither of these effects is explicitly considered by the model here. For the physical effect, a qualitative study can be presented based on the results obtained in Section 3. It can be deduced from the law of effective stress (eg., Scholz, 2002) that the physical effect of the fluid only affects the force $F_N$ in the y-direction. Combining with Eq. (6), the physical effect of the fluid will lead to a decrease in $F_N$ and therefore to a decrease in friction. This in turn will lead to two consequences: the system will remain quasi-static at lower energies state, or the possibility of slope failure will increase.

For the chemical effect of the fluids, it is more complicated and it is usually intertwined with the physical effect. We will not discuss it here. It should be pointed out that we have already subsumed the effects of fluids in the characteristic parameters $\alpha$, $\beta$, and $\gamma$. Because these characteristic parameters are obtained from the actual monitored displacements or velocities with the functional form of Eq. (8) or (10), these actual monitored data has already included the effect of the fluids.

## 5.3 Other applications

The spring-block friction model for landslides in Fig. 2 also can be used to explain the following two observations: (1) The stick-slip of some landslides; (2) The scaling relationship between the numbers of the landslides (or probability) and their volumes.

For simplicity, the following assumptions are made: (1) The static friction coefficient $\mu_s$ and the kinetic friction coefficient $\mu_d$ are constants; (2) $\mu_s$ falls almost instantly to $\mu_d$ when the slipping of the block occurs like that in the classic physics.

Thereby, the block is initially at rest and the spring is pulled with the velocity $v_0$. The spring force increases until it reaches the force $\mu_s F_{gy} - F_{gx}$ at the time

$$t_0 = \frac{\mu_s F_{gy} - F_{gx}}{c v_0} \tag{13}$$

where $F_{gx} = mg \sin \xi$, $F_{gy} = mg \cos \xi$, $\xi$ is the inclination of the slope.

At this moment $t_0$, the block begins to slip and at the same time $\mu_s$ falls to $\mu_d$. The equation of motion of the block and the initial conditions are,

$$\begin{cases} m \frac{d^2 x}{dt^2} = c(v_0 t - x) + F_{gx} - \mu_d F_{gy} \\ x|_{t=t_0} = 0 \\ \frac{dx}{dt}|_{t=t_0} = 0 \end{cases} \tag{14}$$

The solution to (14) is,

$$x = A \sin(\omega t + \varphi) + v_0 t - \frac{\mu_d F_{gy} - F_{gx}}{c} \tag{15}$$

where

$$\begin{cases} A = \frac{1}{\omega} \sqrt{v_0^2 + \left[\omega F_{gy} \frac{\mu_s - \mu_d}{c}\right]^2} \\ \varphi = \arctan \left( \dfrac{v_0 \sin \frac{\omega(\mu_s F_{gy} - F_{gx})}{c v_0} - \omega F_{gy} \frac{\mu_s - \mu_d}{c} \cos \frac{\omega(\mu_s F_{gy} - F_{gx})}{c v_0}}{-v_0 \cos \frac{\omega(\mu_s F_{gy} - F_{gx})}{c v_0} - \omega F_{gy} \frac{\mu_s - \mu_d}{c} \sin \frac{\omega(\mu_s F_{gy} - F_{gx})}{c v_0}} \right) \\ \omega = \sqrt{\frac{c}{m}} \end{cases}$$

The block will come to rest when

$$\frac{dx}{dt} = A\omega \cos(\omega t + \varphi) + v_0 = 0 \tag{16}$$

The corresponding acceleration is,

$$\begin{aligned} \frac{d^2 x}{dt^2} &= -A\omega^2 \sin(\omega t + \varphi) \\ &= -\frac{F_{gy}(\mu_s - \mu_d)}{m} \end{aligned} \tag{17}$$

The force acting on the block is $-F_{gy}(\mu_s - \mu_d)$, and the force of the spring is

$$F_{spring} = -\mu_s F_{gy} + 2\mu_d F_{gy} - F_{gx} \qquad (18)$$
$$= -(\mu_s F_{gy} - F_{gx}) + 2(\mu_d F_{gy} - F_{gx})$$

Because $\mu_d F_{gy} < \mu_s F_{gy}$, $F_{spring} < \mu_s F_{gy} - F_{gx}$. The block will stick until the spring force once again reaches $\mu_s F_{gy} - F_{gx}$. This stick-slip motion of the block will repeat, and the displacement of the block increases with time in a step shape, as shown in Fig. 6. Fig. 6 illustrates that the normalized displacement measurements vary with time for two points A3 and B3 of the Xintan landslide, China. It can be found that the normalized displacement of A3 increases step by step from normalized time 0.2 to 1; While this phenomenon occurs unobviously from 0.5 to 1 for B3.

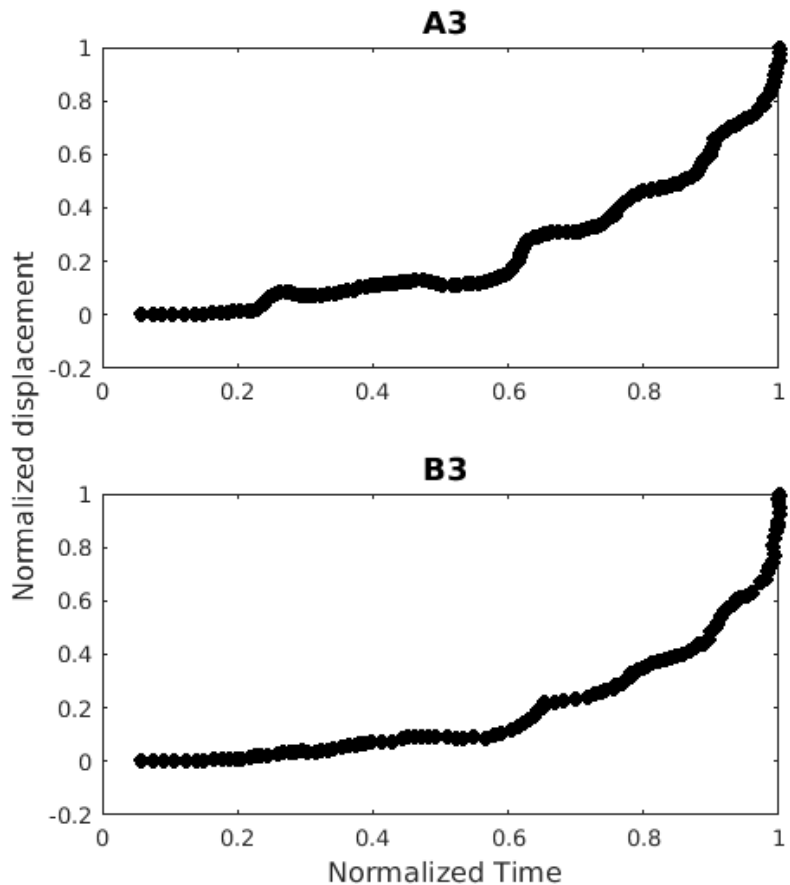

Fig. 6 Normalized displacement measurements vary with time for two points of the Xintan landslide (black dots, digitized from Zhang et al.,. (2005)).

Even more, the displacement of slip for small $v_0$ is,

$$\Delta x = 2F_{gy} \frac{\mu_s - \mu_d}{c} \qquad (19)$$

In the field, there is no individual mass and discrete spring element. However, a simple estimation can be made based on the spring-block friction model. Similar to Popov (2017), we assume that a landslide has a characteristic length $L$; The stiffness with such dimensions is on the order of magnitude $c \approx GL$ ($G$ is the shear moduli). So according to (19) the displacement during a slip event (landslide) is,

$$\Delta x \approx 2F_{gy}\frac{\mu_s - \mu_d}{GL} \approx \frac{2\sigma_N L}{G}(\mu_s - \mu_d) \quad (20)$$

where $\sigma_N = F_{gy}/L^2$ is the normal stress.

Now we consider a region with apparent volume $\bar{V} = \bar{A} \times \bar{L}$ ($\bar{L} >> \Delta x$, $\bar{A}$ is the apparent area, $\bar{L}$ is the apparent height). If only the landslides with characteristic $L$ were possible, then there would be $N$ landslides in this region,

$$N \approx \frac{\bar{A}\bar{L}}{L^2 \Delta x} \approx \frac{G\bar{V}}{2\sigma_N(\mu_s - \mu_d)L^3} \quad (21)$$

Expression (21) shows the frequency of a landslide with a given order of magnitude of the slip displacement is proportional to $L^{-3} = V^{-1}$. However, there is no characteristic length in the real field. We further assume that displacements of different length $L$ can take place with the same probability. Using $\phi(V)$ for the probability density of a landslide, we have,

$$N \propto \phi(V)V \propto V^{-1} \quad (22)$$

Expression (22) infers that $\phi(V) \propto V^{-2}$. Therefore, the probability $\Phi(V)$ of a landslide with volume larger than $V$ is equal to

$$\Phi(V) = \int_V^\infty \phi(V)\mathrm{d}V \propto \int_V^\infty V^{-2}\mathrm{d}V = V^{-1} \quad (23)$$

Fig.7 shows an example for (22) and (23), in which 1579 landslides are triggered by the magnitude M=7.9 Denali earthquake (Alaska, USA) on 3 November 2002. Fig. 7 suggests that the probability of large landslides is small, while the probability of small landslides is large under the same conditions. It should be pointed out the factor "-1" in (22) and (23) could be other values, as the spring block friction model here is too simple to reflect the real landslide.

## 6. Conclusions

With the quasi-static assumption, a simple spring-block friction model using Dieterich-Ruina friction law is developed. By this model, the displacement or velocity of the landslide can be determined by a 3-parameter algebraic equation. Even only the function form is used to study the monitoring displacement or velocity data, it has clear physical basis, which is much more strictly than those methods based on empirical guess.

Applications to the landslides of Vainot and Maoxian show that the kinematic process before the landslide failure can be well represented by the spring-block friction model. Although the estimated failure time of landslide is delayed, it is possible to warn the landslide instability, if with good enough and representative sites, and if with broad, accurate enough monitoring data of displacement or velocity. With the increasing of the amount of high-precision satellite data and the improvement in the data processing algorithms, there is an increasing possibility.

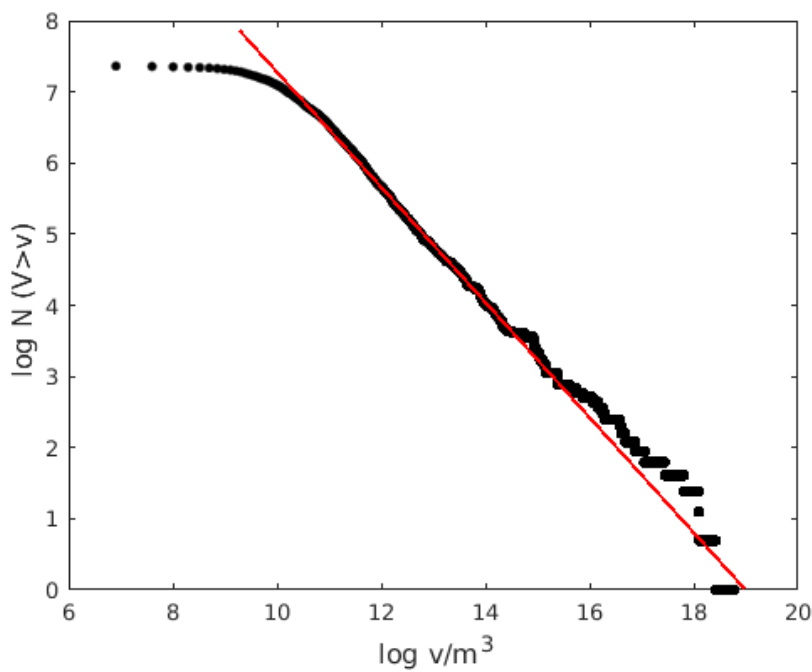

Fig. 7 The number of landslides $N(V > v)$ with a magnitude larger than $v$. The line represents (22) or (23), $\log(N) \propto -b \log v$, with $b = -0.81$. In the figure, 1579 landslides are triggered by an earthquake. Data is from Seal et al. (2022).

## References

Bhandari, R. K., 1988. Some lessons in the investigation and field monitoring of landslides, in Proceedings of the 5th International Symposium on Landslides, vol. 3, edited by C. Bonnard, pp. 1435 – 1457, A. A. Balkema, Brookfield, Vt.

Carlà,T., Intrieri, E., Di Traglia F., Nolesini, T., Gigli, G., Casagli, N., 2017. Guidelines on the use of inverse velocity method as a tool for setting alarm thresholds and forecasting landslides and structure collapses, Landslides, 14, 517–534 Doi: 10.1007 s10346-016-0731-5.

Dieterich, J. H., 1992. Earthquake nucleation on faults with rate and state dependent strength, Tectonophysics, 211, 115-134.

Genevois, R., Ghirotti, M., 2005. The 1963 Vaiont Landslide, Giornale di Geologia Applicata, 1, 41-52, doi: 10.1474/GGA.2005-01.0-05.0005.

Helmstetter, A., Sornette, D., Grasso J.-R., Andersen, J. V., Gluzman, S., and Pisarenko V. 2004. Slider block friction model for landslides: Application to Vaiont and La Clapie`re landslides, J. Geophys. Res., 109, B02409, doi:10.1029/2002JB002160.

Intrieri, E., Raspini, F., Fumagalli, A., Lu, P., Conte, S. D., Farina, P., Allievi, J., Ferretti, A., Casagli, N., 2017. The Maoxian landslide as seen from space: detecting precursors of failure with Sentinel-1 data, Landslides, doi: 10.1007/s10346-017-0915-7.

Kennedy, B. A., and Niermeyer, K. E., 1971. Slope monitoring systems used in the prediction of a major slope failure at the Chuquicamata mine, Chile, in Proceedings on Planning Open Pit Mines, pp. 215– 225, A. A. Balkema, Brookfield, Vt.

Kilburn, C. R. J., and Petley, D. N., 2003. Forecasting giant, catastrophic slop collapse: Lessons from Vaiont, northern Italy, Geomorphology, 54, 21-32.

Popov, V. L., Contact mechanics and friction: physical principles and applications (2nd edition), Berlin: Springer Berlin Heidelberg, 2017.

Popov, V. L., Grzemba, B., Starcevic, J., Popov M., 2012, Rate and state dependent friction laws and the prediction of earthquakes: What can we learn from laboratory models? Tectonophysics, 532-535, 291-300.

Rose, N. D., and Hungr, O., 2007. Forecasting potential rock slope failure in open pit mines using the inverse-velocity method. Int J Rock Mech Min Sci 44:308-320.

Ruina, A., 1983. Slip instability and state variable friction laws, J. Geophys. Res., 88, 10,359–10,370.

Saito, M., 1969. Forecasting time of slope failure by tertiary creep, Proc. Int. Conf. Soil Mech. Found. Eng., 7th, vol. 2, 677–683.

Saito, M., and Uezawa H.， 1961. Failure of soil due to creep, Proc. Int. Conf. Soil Mech. Found. Eng., 6th, vol. 1, 315-318.

Seal, D.M., Jessee, A.N., Hamburger, M.W., Dills, C.W., Allstadt, K.E., 2022, Comprehensive Global Database of Earthquake-Induced Landslide Events and Their Impacts (ver. 2.0, February 2022): U.S. Geological Survey data release, https://doi.org/10.5066/P9RG3MBE.

Scholz, C, H., 2002. The mechanics of earthquakes and faulting, Cambridge University press. pp. 75

Voight, B., 1988. Materials science laws applied to time forecast of slope failure, in Proceedings of the 5th International Symposium on Landslides, vol. 3, edited by C. Bonnard, pp. 1471–1472, A. A. Balkema, Brookfield, Vt.

Wang, Y. P., Xu, Q., Zheng, G., Zheng, H. J., 2015. A rheology experimental investigation on early warning model for landslide based on inverse-velocity method. Rock and soil mechanics, 36 (6): 1606-1614. Doi: 10.16285/j.rsm2015.06.001.

Zeng, Q. L., Wei, R. Q., Xue, X. Y., Zhou, Y. Z., Yin, Q. F., 2018. Characteristics and geohazard mechanism of super-large Xinmo rock avalanche-debris flow in Diexi，Sichuan province, Journal of engineering geology, 1: 193-206,doi：10.13544/j.cnki.jeg.2018.01.021

Zhang, W. J., Chen, Y. M., He K. Q., 2005. Application of loading/unloading response ratio theory to friction of accumulative landslide. Journal of natural disasters, 14 (5): 79-83.